\documentclass[fleqn,usenatbib]{mnras}
\usepackage[english]{babel}

\usepackage[T1]{fontenc}
\usepackage{ae,aecompl}

\usepackage{fancyhdr}
\usepackage{amsfonts}
\usepackage{amsmath}
\usepackage{amssymb}
\usepackage{multicol}
\usepackage{layout}
\usepackage{graphicx}
\usepackage{multirow}

\usepackage{color}
\usepackage{multirow}

\usepackage{times}
\usepackage{natbib}

\newif\ifAMStwofonts
\AMStwofontstrue

\graphicspath{{./fig/}}

\title[SCM in f(T) cosmology]{Spherical collapse model and cluster number counts in power law $f(T)$ gravity}

\author[M. Malekjani et~al.]{M. Malekjani$^{1}$  \thanks{malekjani@basu.ac.ir},
{S. Basilakos $^{2}$}, {N. Heidari $^{1}$}
 \\
$^1$ Department of Physics, Bu Ali Sina University, Hamedan 65178, Iran\\
$^2$ Academy of Athens, Research Center for Astronomy and Applied
Mathematics, Soranou Efessiou 4, 11-527 Athens, Greece}

\date{Accepted ?, Received ?; in original form \today}

\pagerange{\pageref{firstpage}--\pageref{lastpage}} \pubyear{2015}

\begin{document}

\label{firstpage}

\maketitle
\begin{abstract}
We study the spherical collapse model (SCM) in the framework of
spatially flat power law $f(T) \propto (-T)^{b}$  gravity model. We
find that the linear and non-linear growth of spherical
overdensities of this particular $f(T)$ model are affected by the
power-law parameter $b$. Finally, we compute the predicted number
counts of virialized haloes in order to distinguish the current
$f(T)$ model from the expectations of the concordance $\Lambda$
cosmology. Specifically, the present analysis suggests that the
$f(T)$ gravity model with positive (negative) $b$ predicts more
(less) virialized objects with respect to those of $\Lambda$CDM.
\end{abstract}


\section{Introduction}

The idea of the accelerated expansion of the universe is supported
by several independent cosmological experiments including those of
supernova type Ia \citep{Riess1998,Perlmutter1999,Kowalski2008},
cosmic microwave background (CMB)
\citep{Komatsu2009,Komatsu2011,Jarosik:2010iu,Ade:2015yua},
large scale structure and baryonic acoustic oscillation
\citep{Percival2010,Tegmark:2003ud,Cole:2005sx,Eisenstein:2005su,Reid:2012sw,Blake:2011rj},
high redshift galaxies \citep{Alcaniz:2003qy}, high redshift galaxy
clusters \citep{Allen:2004cd,Wang1998} and weak gravitational
lensing \citep{Benjamin:2007ys,Amendola:2007rr,Fu:2007qq}. Cosmic
acceleration can well be interpreted in the framework of general
relativity (GR) by invoking the dark energy (DE) component in the
total energy budget of the universe. Although, the earliest and
simplest candidate for DE is the traditional cosmological constant
$\Lambda$ with constant equation of state (EoS) parameter
$w_{\Lambda}=-1$ \citep{Peebles2003}, the well known issues which
are associated with the fine-tuning and cosmic coincidence problems,
\citep{Weinberg1989,Sahni:1999gb,Carroll2001,Padmanabhan2003,Copeland:2006wr}
has led the scientific community to propose a large family of
dynamical DE models (quintessence
\citep{Caldwell:1997ii,Erickson:2001bq}, phantom
\citep{Caldwell2002}, k-essence \citep{Armendariz2001}, tachyon
\citep{Padmanabhan2002}, quintom \citep{Elizalde:2004mq}, Chaplygin
gas \citep{Kamenshchik:2001cp} and generalized Chaplygin gas
\citep{Bento:2002ps} etc) in which $w_{\rm de}\neq-1$.

On the other hand, one can consider that cosmic acceleration
reflects on the physics of gravity on cosmological scales. Indeed,
modifying the Einstein-Hilbert action and using the
Friedmann-Robertson-Walker (FRW) spacetime as a background metric
one can obtain the modified Friedmann's equations which can be used
in order to understand the accelerated expansion of the universe. As
an example, one of the most popular modified gravity models is the
$f(R)$ scenario in which we allow the Lagrangian of the modified
Einstein-Hilbert action to be a function of the Ricci scalar $R$
\citep{Capozziello:2007ec,Nojiri:2010wj,Sotiriou:2008rp}.
Alternatively, among the large group of extended theories of gravity
the so-called $f(T)$ gravity plays an important role in this kind of
studies. This theory is based on the old definition of the
teleparallel equivalent of general relativity (TEGR), first
introduced by \cite{Einstein29} \citep[see
also][]{Hayashi:1979qx,Maluf:1994ji}. Here, instead of
using the curvature defined through the Levi-Civita connection one
can assume an alternative approach based on torsion $T$ via the
Weitzenb\"{o}ck connection in order to extract the torsion scalar
\citep{Hayashi:1979qx}. Inspired by the methodology of $f(R)$
gravity, a natural extension of TEGR is the theory of $f(T)$ gravity
in which we assume that the Lagrangian of the modified
Einstein-Hilbert action is a function of $T$
\citep{Ferraro:2006jd,Linder:2010py}. It is worth noting that in
$f(T)$ gravity we have second-order field equations while in $f(R)$
gravity we deal with fourth-order field equations which may lead to
pathologies as discussed in the work of
\cite{Capozziello:2009pi,Capozziello:2009mq}. In the literature,
there are plenty of papers available that study the cosmological
properties of different $f(T)$ models. In particular, the background
history and the cosmic acceleration can be found in Refs.
\citep{Bengochea:2008gz,Linder:2010py,Myrzakulov:2010vz,Dent:2011zz,Zhang:2011qp,Capozziello:2011hj,Geng:2011aj,Bamba:2012vg}.
The dynamical aspects and the cosmological constraints of the $f(T)$
models have been investigated in Refs.
\citep{Wu:2010xk,Wu:2010av,Dent:2011zz,Bamba:2010wb,Capozziello:2011hj,Geng:2011aj,Wei:2011yr,Karami:2012if,Bamba:2012vg}
and in Refs.
\citep{Wu:2010mn,Nunes:2016qyp,Saez-Gomez:2016wxb,Geng:2011ka,Wei:2011mq,Geng:2011ka,Cardone:2012xq,Iorio:2015rla}.
Also the connection between $f(T)$ and scalar field theory can be
found in \citep{Yerzhanov:2010vu,Chen:2014qsa,Sharif:2013wga}.
Lastly, at the perturbation level we refer the reader the works of
Refs.
\citep{Chen:2010va,Zheng:2010am,Wu:2011kh,Li:2011wu,Wu:2012hs,Izumi:2012qj,Geng:2012vn,Basilakos:2016xob}.

Is is well known that in order to distinguish modified gravity
models from scalar field DE models we need to study the growth of
matter perturbations in linear and non-linear regimes. Specifically,
the growth index $\gamma$ of linear matter fluctuations \cite[first
introduced by][]{Peebles1993} in $f(T)$ gravity is investigated in
\cite{Zheng:2010am,Basilakos:2016xob}. \cite{Basilakos:2016xob}
found that the asymptotic form of the power-law $f(T)$ model is
given by $\gamma \approx \frac{6}{11-6b}$ which naturally extends
that of the $\Lambda$CDM model, $\gamma_{\Lambda}\approx 6/11$.

The spherical collapse model (herafter SCM), first introduced by
\cite{Gunn1972}, is a simple analytical approach to study the
evolution of the growth of matter fluctuations in the non-linear
regime. Notice, that the scales of SCM are much smaller than the
Hubble radius and the velocities are non-relativistic. The central
idea of the SCM  is based on the fact that due to self-gravity,
we expect that the spherical overdensities expand with slower rate
than the Hubble expansion. Therefore, at a certain redshift the
over-dense region completely decouple from the background expansion
(reaching to a maximum radius) and it starts to 'turn around'. This
redshift is the so-called turn around redshift, $z_{\rm ta}$. After
$z_{\rm ta}$, the spherical region collapses due to self gravity and
finally it reaches the steady state virial radius at a certain
redshift $z_{\rm vir}$. In the framework of General Relativity (GR),
the SCM has been investigated in several independent works
\citep{Fillmore1984,Bertschinger1985,Hoffman1985,
Ryden1987,AvilaReese:1997us,Subramanian2000,
Ascasibar2004,Williams2004,Mehrabi:2016exz}. Also, the SCM has been extended for
various cosmological models, including those of DE
\citep{Mota2004,Maor2005,Basilakos:2006us,Basilakos2009,Li2009,Pace2010,Wintergerst2010,Basse2011,
Pace2012,Naderi2015,Abramo2007,Abramo2009a,Malekjani:2015pza},
scalar-tensor and modified gravity
\citep{Schaefer:2007nf,Pace:2013pea,Nazari-Pooya:2016bra,Fan:2015lta}.
We would like to stress that the general formalism of SCM 
can be used in the case where  
Birkhoff's theorem is valid. 
As an example, $f(R)$ gravity models which are 
based on metric formalism can not accommodate 
Birkhoff's theorem, while in the case of Palatini formalism this theorem holds
\citep{Sotiriou:2008rp,Capozziello:2007ms,Faraoni:2010rt}. 
In $f(T)$ gravity, it has been shown that the Birkhoff's 
theorem is valid \citep{Meng:2011ne} and thus one can 
extend the SCM in the context of $f(T)$ models.

In the present article, we attempt to study the non-linear growth of
matter overdensities and the corresponding number counts of the
power law $f(T)$ model
\citep{Linder:2010py,Ferraro:2006jd,Ferraro:2008ey} \cite[see
also][and references therein]{Cai:2015emx}. To the best of our
knowledge, we are unaware of any previous investigation regarding
the SCM in $f(T)$ gravity and thus we believe that the current
analysis can be of theoretical interest. Notice, that the growth of
matter perturbations in the linear regime has been investigated in
\citep{Chen:2010va,Basilakos:2016xob}.

We organize our paper as follows. In section \ref{ft model}, we
briefly present the basic cosmological properties of $f(T)$ gravity
and then we focus on the power-law model. In section \ref{grwoth} we
study the growth of matter fluctuations in the linear and non-linear
(SCM) regimes respectively. In section \ref{mass function}, we
compute the predicted mass function and the number counts of the
power-law $f(T)$ model and we discuss the differences from the
concordance $\Lambda$ cosmology. Finally, we provide our conclusions
in section \ref{conclusion}.

\section{background history in power law $f(T)$ model}
\label{ft model}
 In this section we briefly present the main points
of the $f(T)$ gravity \citep[see also][and references
therein]{Basilakos:2016xob}. In particular, the action in the case
of $f(T)$ gravity is given by
\begin{eqnarray}
\label{action1}
 I = \frac{1}{16\pi G_N }\int d^4x e
\left[T+f(T)+L_m+L_r\right],
\end{eqnarray}
where $L_{\rm m}$ and $L_{\rm r}$ are the matter and radiation
Lagrangians respectively. Notice, that $e =
\text{det}(e_{\mu}^A)=\sqrt{-g}$ and ${\mathbf{e}_A(x^\mu)}$ are the
vierbein fields. In this context, the gravitational field is
expressed in terms of torsion tensor which produces (after the
necessary contractions) the torsion scalar $T$
\citep{Hayashi:1979qx}.

Varying the above action with respect to the vierbeins the modified
Einstein's field equations are
\begin{eqnarray}\label{eom}
&&e^{-1}\partial_{\mu}(ee_A^{\rho}S_{\rho}{}^{\mu\nu})[1+f_{T}]
 +
e_A^{\rho}S_{\rho}{}^{\mu\nu}\partial_{\mu}({T})f_{TT}\ \ \ \ \  \ \
\ \  \ \
\ \ \nonumber\\
&& \ \ \ \
-[1+f_{T}]e_{A}^{\lambda}T^{\rho}{}_{\mu\lambda}S_{\rho}{}^{\nu\mu}+\frac{1}{4}
e_ { A } ^ { \nu
}[T+f({T})] \nonumber \\
&&= 4\pi Ge_{A}^{\rho}\overset {\mathbf{em}}T_{\rho}{}^{\nu},
\end{eqnarray}
where $f_{T}=\partial f/\partial T$, $f_{TT}=\partial^{2} f/\partial
T^{2}$, and $\overset{\mathbf{em}}{T}_{\rho}{}^{\nu}$ represents the
standard energy-momentum tensor. Considering the description of
perfect fluids the energy momentum tensor takes the form
\begin{eqnarray}\label{prefect1}
\overset{\mathbf{em}}T_{\rm \mu \nu}{}=Pg_{\rm
\mu\nu}-(\rho+P)u_{\rm \mu}u_{\rm \nu},
\end{eqnarray}
where $u^{\rm\mu}$ is the fluid four-velocity, $\rho=\rho_{\rm m}+\rho_{\rm r}$
is the total pressure and $P=P_{\rm m}+P_{\rm r}$ is the total pressure with
$(P_{\rm m},P_{\rm r})=(0,\rho_{\rm r}/3)$.
Of course $\rho_{\rm m}$ ($\rho_{\rm r}$) and $P_{\rm m}$
($P_{\rm r}$) denotes the energy density and pressure
of the non-relativistic matter (radiation) respectively.
In the matter dominated
era and prior to the present time we can neglect the radiation
component from the cosmic expansion. Through out the
current work
we consider the usual form of the vierbiens
\begin{equation}
\label{weproudlyuse} e_{\mu}^A={\rm diag}(1,a,a,a),
\end{equation}
which leads to a flat FRW metric
\begin{equation}
ds^2= dt^2-a^2(t)\,\delta_{ij} dx^i dx^j,
\end{equation}
where $a(t)$ is the scale factor of the universe.
Now, inserting the aforementioned vierbeins and the
energy momentum tensor into the field
equations (\ref{eom}) we can provide the modified Friedmann
equations 
\begin{eqnarray}\label{background1}
&&H^2= \frac{8\pi G_N}{3}(\rho_m+\rho_r)
-\frac{f}{6}+\frac{Tf_T}{3},\\\label{background2}
&&\dot{H}=-\frac{4\pi G_N(\rho_m+P_m+\rho_r+P_r)}{1+f_{T}+2Tf_{TT}},
\end{eqnarray}
where the overdot represents the derivative with respect to cosmic
time $t$ and $H\equiv\dot{a}/a$ is the Hubble parameter. The Hubble
parameter $H$ in $f(T)$ gravity is given in terms of $T$ via the
relation
\begin{eqnarray}
\label{TH2} T=-6H^2 \;.
\end{eqnarray}
From equation (\ref{TH2}), it is easy to prove that the
dimensionless Hubble parameter is given by
\begin{eqnarray}
\label{TH3} E^{2}(a)\equiv\frac{H^2(a)}
{H^2_{0}}=\frac{T(a)}{T_{0}},
\end{eqnarray}
which gives
\begin{equation}
\label{TDE} \frac{d{\rm ln}T}{d{\rm ln}a}=2T_{0}E(a)\frac{d{\rm
ln}E}{d{\rm ln}a}\;,
\end{equation}
where $H_{0}$ is the Hubble constant and $T_0\equiv-6H_{0}^{2}$.
From equations (\ref{background1} \& \ref{background2}) we can
obtain the energy density and the pressure of the effective DE
component as follows \citep{Linder:2010py}
\begin{eqnarray}
&&\rho_{\rm de}\equiv\frac{3}{8\pi
G_N}\left[-\frac{f}{6}+\frac{Tf_T}{3}\right], \label{rhoDDE}\\
\label{pDE} &&P_{\rm de}\equiv\frac{1}{16\pi G_N}\left[\frac{f-f_{T}
T +2T^2f_{TT}}{1+f_{T}+2Tf_{TT}}\right].
\end{eqnarray}
The corresponding effective equation of state (EoS) parameter
is written as
\begin{eqnarray}
\label{wfT}
 w_{\rm de}=\frac{P_{\rm de}}{\rho_{\rm de}}=
-1-\frac{1}{3}\frac{d{\rm ln}T}{d{\rm ln}a}
\frac{f_{T}+2Tf_{TT}}{[(f/T)-2f_{T}]}.
\end{eqnarray}
Utilizing equation (\ref{background1}) and the nominal relations
$\rho_{m}=\rho_{m0}a^{-3}$ and $\rho_{r}=\rho_{r0}a^{-4}$ we compute
the dimensionless Hubble parameter
\begin{eqnarray}\label{Mod1Ez}
E^2(a)=\Omega_{m0}a^{-3}+\Omega_{r0}a^{-4}+\Omega_{F0} X(a),
\end{eqnarray}
where $\Omega_{i0}=\frac{8\pi G
\rho_{\rm i0}}{3H_0^2}$,
$\Omega_{F0}=1-\Omega_{m0}-\Omega_{r0}$ and the
function $X(a)$ is given by
\begin{equation}
\label{distortparam}
 X(a)=\frac{1}{T_0\Omega_{F0}}\left(f-2Tf_T\right).
\end{equation}
Evidently, the Hubble expansion in $f(T)$ cosmology is
affected by the extra term $\Omega_{F0} X(a)$
which is given in terms of functional form of
$f(T)$, as indicated from equation
(\ref{distortparam}).

For the rest of the paper we focus our analysis on the power law
$f(T)$ pattern \citep{Bengochea:2008gz} in which the form of $f(T)$
is given by
\begin{equation}
\label{Pow} f(T)=\alpha (-T)^{b},
\end{equation}
where $\alpha=(6H_0^2)^{1-b}\frac{\Omega_{F0}}{2b-1}$. Substituting
(\ref{Pow}) into equations (\ref{wfT}) and (\ref{distortparam}) we
can get
\begin{equation}
\label{yLL} X(a,b)=E^{2b}(a,b),
\end{equation}

\begin{equation}\label{TDE1}
w_{\rm de}=-1-\frac{2b}{3}\frac{d{\rm ln}E}{d{\rm
ln}a}
\end{equation}
and inserting (\ref{yLL}) into equation (\ref{Mod1Ez}) we arrive at
\begin{eqnarray}
\label{Mod1Ezz} E^2(a,b)=\Omega_{\rm m0}a^{-3}+\Omega_{\rm
r0}a^{-4}+\Omega_{\rm F0} E^{2b}(a,b) \;.
\end{eqnarray}
As expected, for $f(T)={\rm const.}$ the above cosmological
quantities boil down to those of $\Lambda$CDM ($\Omega_{\rm
\Lambda,0}\equiv \Omega_{\rm F0}$). Theoretically, it has been found
that in order to treat the accelerated expansion of the universe the
free parameter $b$ needs to satisfy the condition $b\ll1$
\citep{Linder:2010py,Nesseris:2013jea}. Under these circumstances
the $f(T)$ power law model can be viewed as a perturbation around
the $\Lambda$CDM cosmology
\citep{Nesseris:2013jea,Basilakos:2016xob}. Hence, we can perform a
Taylor expansion of $E^2(a,b)$ around $b=0$ as
$$
E^2(a,b)=E^2(a,0)+\left.\frac{dE^2(a,b)}{db}\right|_{b=0} b+...
$$
or
\begin{equation}
E^2(a,b)=E^2_{\Lambda}(a)+\Omega_{F0}\left.\frac{dX(a,b)}{db}\right|_{b=0}
b+...\;, \label{ModapE}
\end{equation}
where for the latter equality we have used Eq.(\ref{distortparam}).
Utilizing equation (\ref{yLL}), we can easily provide a useful
approximate formula of the
dimensionless Hubble parameter 
\cite[see also][]{Basilakos:2016xob}
\begin{equation}
\label{approxM1}
 E^2(a,b)\simeq
E^2_\Lambda(a)+\Omega_{F0}\ln\left[E^2_\Lambda(a)\right]b \;,
\end{equation}
where $E^2_{\rm \Lambda}(a)\equiv E^2(a,0)=\Omega_{\rm
m0}/a^3+\Omega_{\rm r0}/a^4+\Omega_{\rm F0}$. Obviously, the
background evolution of universe depends directly from the
free parameters $b$ and $\Omega_{\rm m0}$. Notice,
that as we have already mentioned above
at late enough times we can neglect the
radiation component from the Hubble parameter which means that
$\Omega_{\rm F0}$ is determined via $\Omega_{\rm F0}=1-\Omega_{\rm
m0}$ for a spatially flat FRW metric.

\begin{figure}
\centering
\includegraphics[width=8cm]{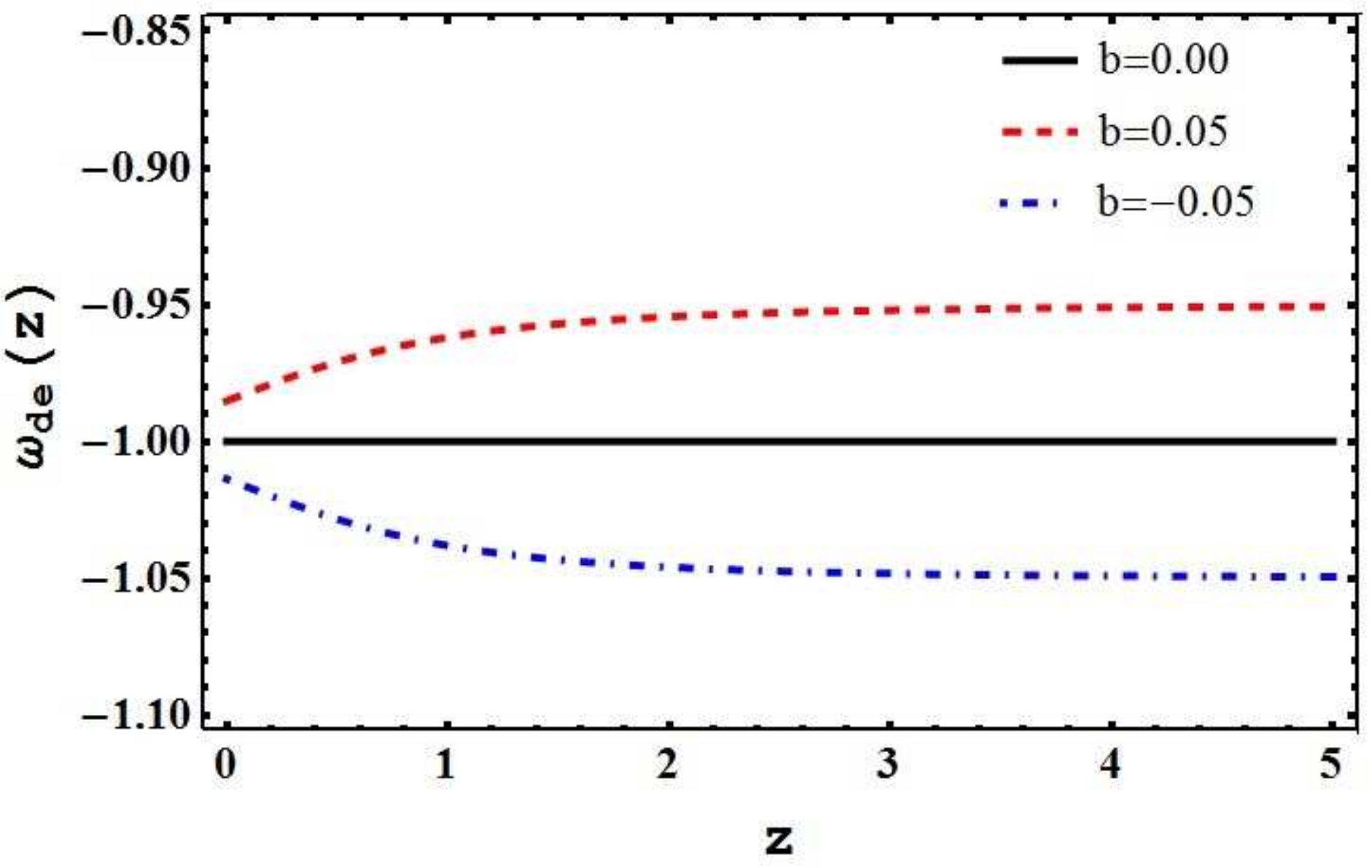}
\includegraphics[width=8cm]{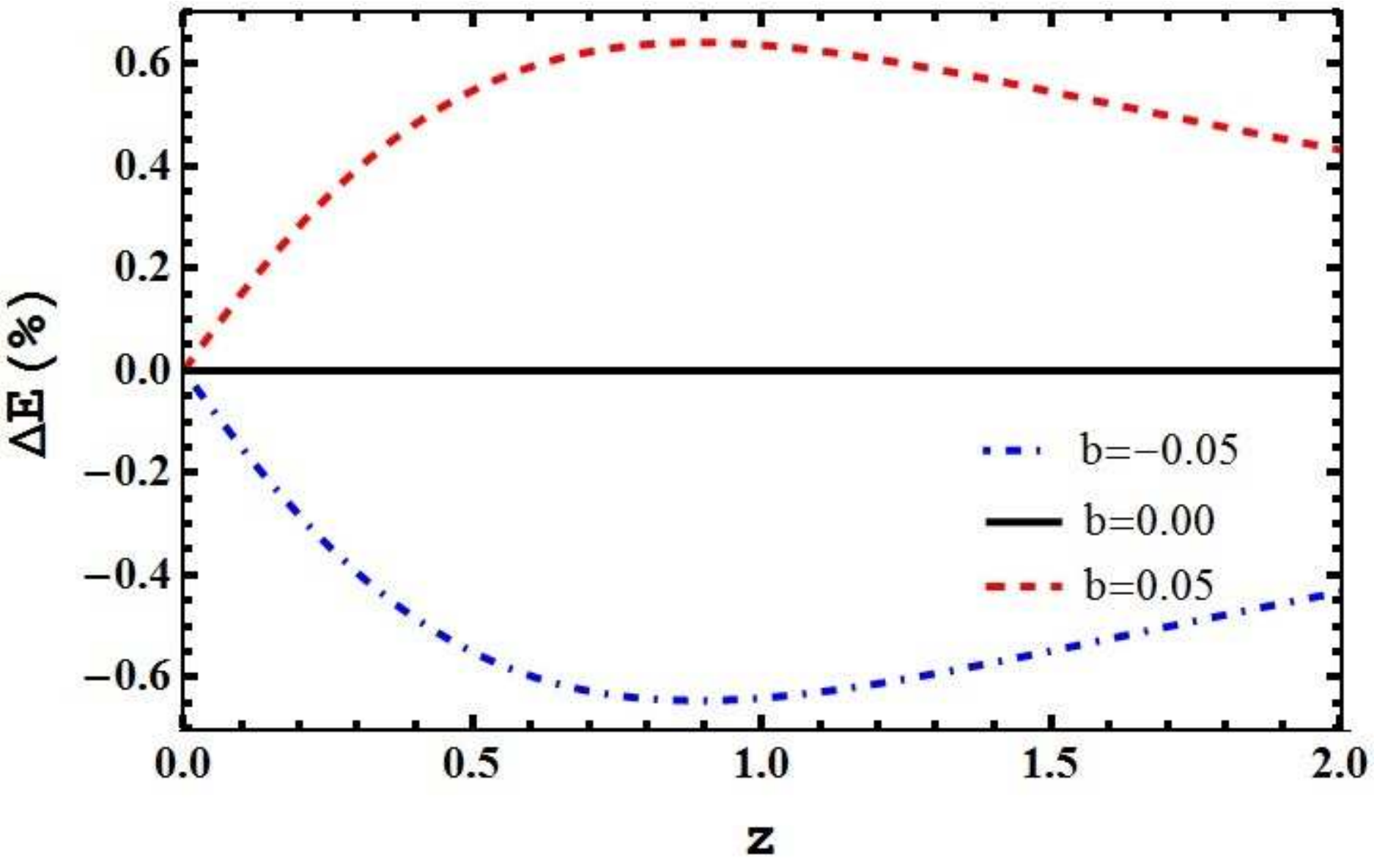}
\includegraphics[width=8cm]{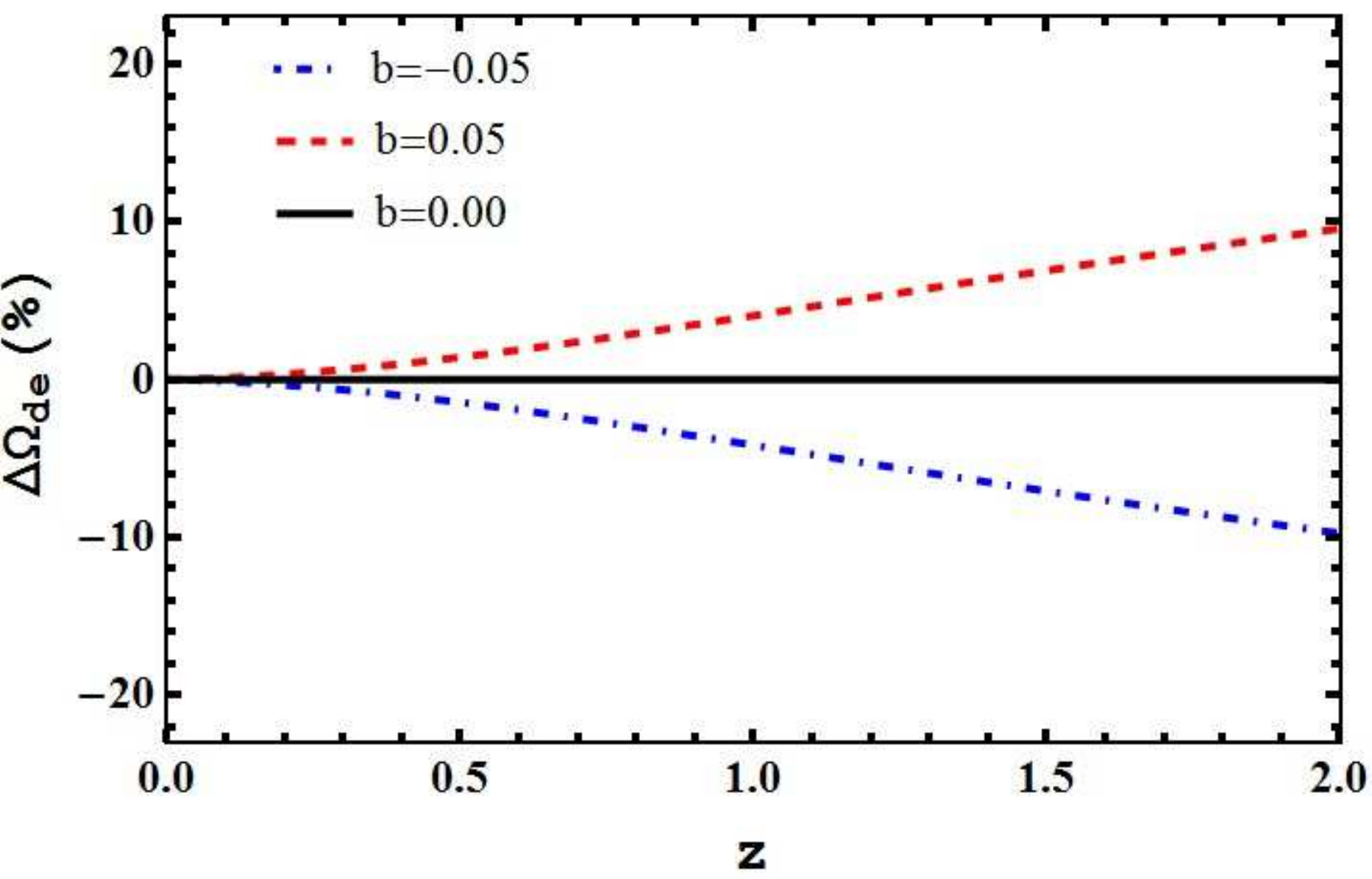}
 \caption{{\it Top panel:} The evolution of the effective EoS parameter
$w_{\rm de}$. {\it Middle panel:}
The fractional difference $\Delta E$
between the power law $f(T)$ model and the reference
$\Lambda$CDM model. {\it Bottom panel:}
The fractional deviation $\Delta \Omega_{\rm d}$ versus redshift.
The curves correspond to the following cosmological models: (i)
$\Lambda$CDM (black solid, $b=0.00$), (ii) $f(T)$ (red dashed line, $b=0.05$)
and (iii) $f(T)$ (blue dotted-dashed line, $b=-0.05$).} \label{fig:background}
\end{figure}

Recently, using the latest observational data that include SNIa
\citep{Union2.1:2012}, BAO \citep{Blake:2011en,Percival2010} and
Planck CMB shift parameter \citep{Shafer:2013pxa} it has been found
that $\Omega_{m0}=0.286\pm 0.012$, $b=-0.081\pm 0.117$
\citep{Basilakos:2016xob}. These results are in agreement (within
$1\sigma$ uncertainties) with those of 
\cite{Nesseris:2013jea} who found $\Omega_{m0}=0.274\pm 0.008$,
$b=-0.017\pm 0.083$. We observe that the above analysis provide a
small and negative value for $b$ but the $1\sigma$ error is quite
large. In order to realize the differences of the power-law $f(T)$
model from the $\Lambda$ cosmology at the expansion level we plot in
Fig.(\ref{fig:background}) the evolution of the EoS parameter $w_{\rm de}(z)$ (top
panel), $\Delta E=\left[(E(a,b)-E_{\rm \Lambda})/E_{\rm
\Lambda}(a)\right]\times 100$ (middle panel) and $\Delta \Omega_{\rm
de}=\left[(\Omega_{\rm de}(a,b)-\Omega_{\rm \Lambda}(a))/\Omega_{\rm
\Lambda}(a)\right]\times 100$ (bottom panel). Notice that the solid,
dashed and dotted-dashed lines correspond to different values of the
$b$ parameter, namely 0, $0.05$ and $-0.05$. Concerning the value of
$\Omega_{\rm m0}$ we have set it to $0.30$ which means that
$\Omega_{\rm F0}=0.70$. Overall, the evolution of the aforementioned
cosmological quantities depends on the model parameter $b$. We
verify that in the case of $b<0$ the effective EoS parameter of the
power law $f(T)$ model remains in the quintessence regime ( $w_{\rm
de}>-1$), while it goes to phantom ($w_{\rm de}<-1$) for $b>0$.
Furthermore, from Fig.(\ref{fig:background}) (see middle and bottom panels) we observe
that in the case of $b>0$ the cosmological quantities $E(z)$ and
$\Omega_{\rm d}(z)$ of the $f(T) \propto (-T)^{b}$ model are large
with respect to those of the reference $\Lambda$CDM model. The
opposite holds for negative values of $b$. Regarding, the Hubble
parameter we find that close to $z\sim 1$ the relative deviation
$\rm \Delta E$ lies in the interval $[-0.6\%,0.6\%]$ for $-0.05\le b
\le 0.05$, while the relative difference $\rm \Delta \Omega_{\rm d}$
can reach up to $\pm 10\%$ at large redshifts $z\sim 2$.

\section{growth of overdensities in $f(T) \propto (-T)^{b}$ gravity}
\label{grwoth}

In this section we explore the growth of matter over-densities
in the $f(T) \propto (-T)^{b}$ model. First, we focus on the
linear perturbation theory and then with the aid
of the SCM we study the non-linear matter fluctuations.

\subsection{linear growth factor}
Let us start with the linear growth of non-relativistic ($P_{m}=0$)
perturbations. In general at the sub-horizon scales
matter perturbations $\delta_{\rm m}$ satisfies the following
differential equation

\begin{equation}
\label{eq:111} \ddot{\delta}_{m}+2 H\dot{\delta}_{m}-4\pi G_{\rm
eff} \rho_{m} \delta_{m}=0 \;,
\end{equation}
where $G_{\rm eff}$ is the effective Newton's parameter and in the
case of $f(T)$ gravity models it takes the form \citep{Zheng:2010am}
\begin{equation}\label{Geff}
G_{\rm eff}=\frac{G_{\rm N}}{1+f_{\rm T}}\;,
\end{equation}
where $G_{\rm N}$ is the Newton's constant. Of course for Einstein's
gravity we have $G_{\rm eff}=G_{\rm N}$. Now, combining equation
(\ref{Pow}) and equation (\ref{Geff}) we obtain
\begin{equation}
G_{\rm eff}=
\frac{G_{\rm N}}{1+
\frac{b\Omega_{F0}}{(1-2b)E^{2(1-b)}}}
\end{equation}
and utilizing a first order Taylor expansion around $b = 0$ we find
\begin{equation}
\label{Geff1}
G_{\rm eff}\simeq G_{\rm N}\left(
1-\frac{\Omega_{F0}}{E^{2}_{\Lambda}(a)}\;b\right)
\end{equation}

Inserting equation (\ref{Geff1}) into equation (\ref{eq:111}) and
changing variables from cosmic time to scale factor ($d/dt=aHd/da$) we find after
some calculations
\begin{equation}
 \delta ''_{\rm m}+\left(\frac{3}{a}+\frac{E^{\prime}(a)}{E(a)}\right)\delta'_{\rm m}
-\frac{3\Omega_{\rm m0}}{2a^5E^{2}(a)}(1-\frac{\Omega_{\rm F0}b}{E^2_{\rm \Lambda}(a)})\delta_{\rm
 m}
 =0 \;,\label{eqc1}
 \end{equation}
where $\delta'_{\rm m}=d \delta_{\rm m}/da$, $\delta ''_{\rm
m}=d^{2} \delta_{\rm m}/da^{2}$ and $E(a)$ is given by equation
(\ref{approxM1}). As expected for $b=0$ the above equation reduces
to that of $\Lambda$CDM presented in \cite[][and references
therein]{Pace2010}.

\begin{figure}
 \centering
 \includegraphics[width=8cm]{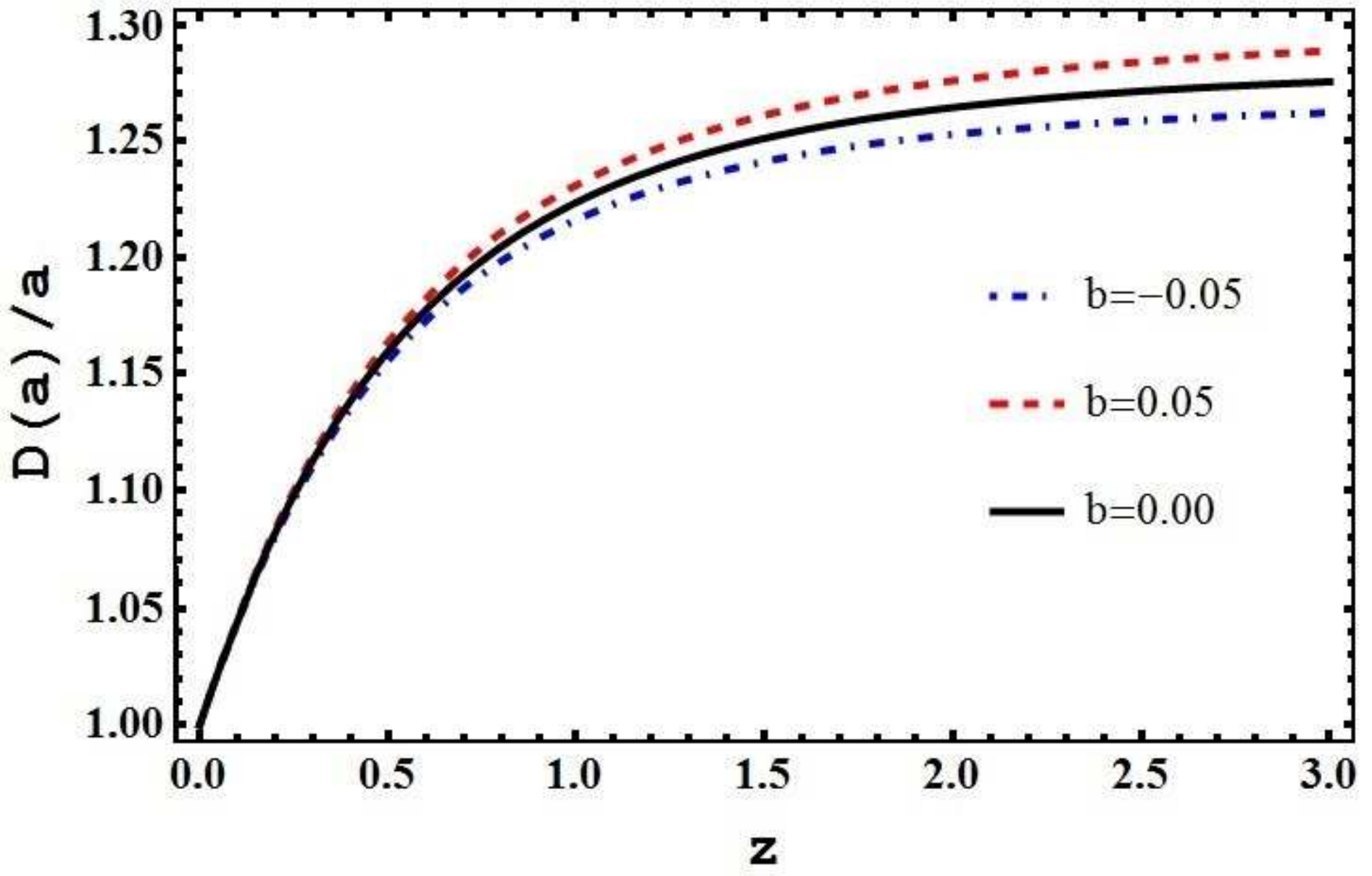}
 \caption{The ratio $D(a)/a$ as a function of $z$.
The style of lines
can be found in the caption of Fig. (\ref{fig:background}).}
 \label{fig:gf}
\end{figure}

Now we numerically integrate equation (\ref{eqc1}) starting from the
initial scale factor $a_{\rm i}=10^{-4}$ till the present epoch
$a=1$. Regarding the initial conditions we adopt the following case:
at $a_{\rm i}=10^{-4}$ we use $\delta_{\rm mi}(a_{\rm i})=1.5 \times
10^{-5}$. Additionally, we also adopt the initial conditions and
$\delta^{\prime}_{\rm mi}=\delta_{\rm mi}/a_i$ which guarantees that
matter perturbations grow in the linear regime \cite[see
also][]{Batista:2013oca,Mehrabi:2015hva,Mehrabi:2015kta,Malekjani:2016edh}. Once the linear matter
overdensity $\delta_{\rm m}$ is found we compute the linear growth
factor scaled to unity at the present time $D(a)=\delta_{\rm
m}(a)/\delta_{\rm m}(a=1)$. In Fig.(\ref{fig:gf}), we show
$D(a)/a$ as a function of redshift ($z=1/a-1$). It is well known
that for the Einstein de-Sitter (EdS) model ($\Omega_{\rm m}=1)$ the
growth factor is proportional to $a$ which implies that $D(a)/a$ is
always equal to unity. For the concordance $\Lambda$ cosmology
($b=0$ black solid curve), the growth factor $D_{\rm \Lambda}(a)/a$
is higher than the EdS model at high redshifts and progressively it
starts fall down at low redshifts. The decrement of the growth
factor at late times shows that the cosmological constant $\Lambda$
dominates the energy budget of the universe and consequently
suppresses the growth of matter overdensities. The opposite is true
at high redshifts, meaning that the effect of cosmological constant
$\Lambda$ on the growth of perturbations is actually negligible and
thus $D_{\rm \Lambda}/a$ reaches a plateau. The above general
behavior holds also for the power-law $f(T)$ model with one
difference namely, for $b=0.05$ (or -0.05) the amplitude of $D(a)/a$
is somewhat larger (or lower) than the $\Lambda$CDM  model at high
redshifts. Specifically, for $z \ge 3$ we find that the relative
difference is $\sim\pm 1\%$. Qualitatively speaking, 
these results are in agreement with those of 
DE models \citep[see][]{Pace2010,Devi:2010qp,Pace:2013pea,Nazari-Pooya:2016bra} .

\subsection{The spherical collapse model}
The spherical collapse model \citep{Gunn1972} is a simple but still
a useful tool utilized to investigate the growth of bound systems in
the universe through gravitational instability\,\citep{Peebles1993}.
It is well known that the main quantities of the SCM, such as the
linear overdensity parameter $\delta_{\rm c}$ and the virial
overdensity $\Delta_{\rm vir}$, are affected by the presence of dark
energy
\citep{Lahav1991,Wang1998,Mota2004,Horellou2005,Wang2005,Abramo2007,Basilakos:2006us,
Pace2010,Pace2012,Batista:2013oca,Pace:2013pea,Pace:2014taa,Malekjani:2015pza,Naderi2015}.
Here our aim is to extent SCM within the $f(T)$ cosmological
scenario, in order to derive the non-linear structure formation in
such models and study the differences with the corresponding
predictions of the usual $\Lambda$CDM cosmology.

Since Birkhoff's theorem holds here, we can start from 
the differential equation that
describes the growth of matter overdensities in the non-linear
regime \citep[see also][]{Pace:2013pea}
\begin{eqnarray}
 \ddot{\delta}_{\rm m}+2H\dot{\delta}_{\rm m}-\frac{4}{3}\frac{\dot{\delta}_{\rm m}^2}{1+\delta_{\rm m}} -
 4\pi G_{\rm eff}\rho_{\rm m}\delta_{\rm m}(1+\delta_{\rm m})=0 \;.\label{matperteq3}
\end{eqnarray}
In the linear regime the above equation reduces to equation
(\ref{eq:111}) as it should. Also, in the case of GR the full
derivation of equation (\ref{eq:111}) can be found in
Ref.\citep{Abramo2007}. It is interesting to mention that the
non-linear matter fluctuations are affected by the law of gravity
via the form of $G_{\rm eff}$. In the case of $f(R)$ gravity we
refer the reader the work of \cite{Schaefer:2007nf}.

\begin{figure}
 \centering
 \includegraphics[width=8cm]{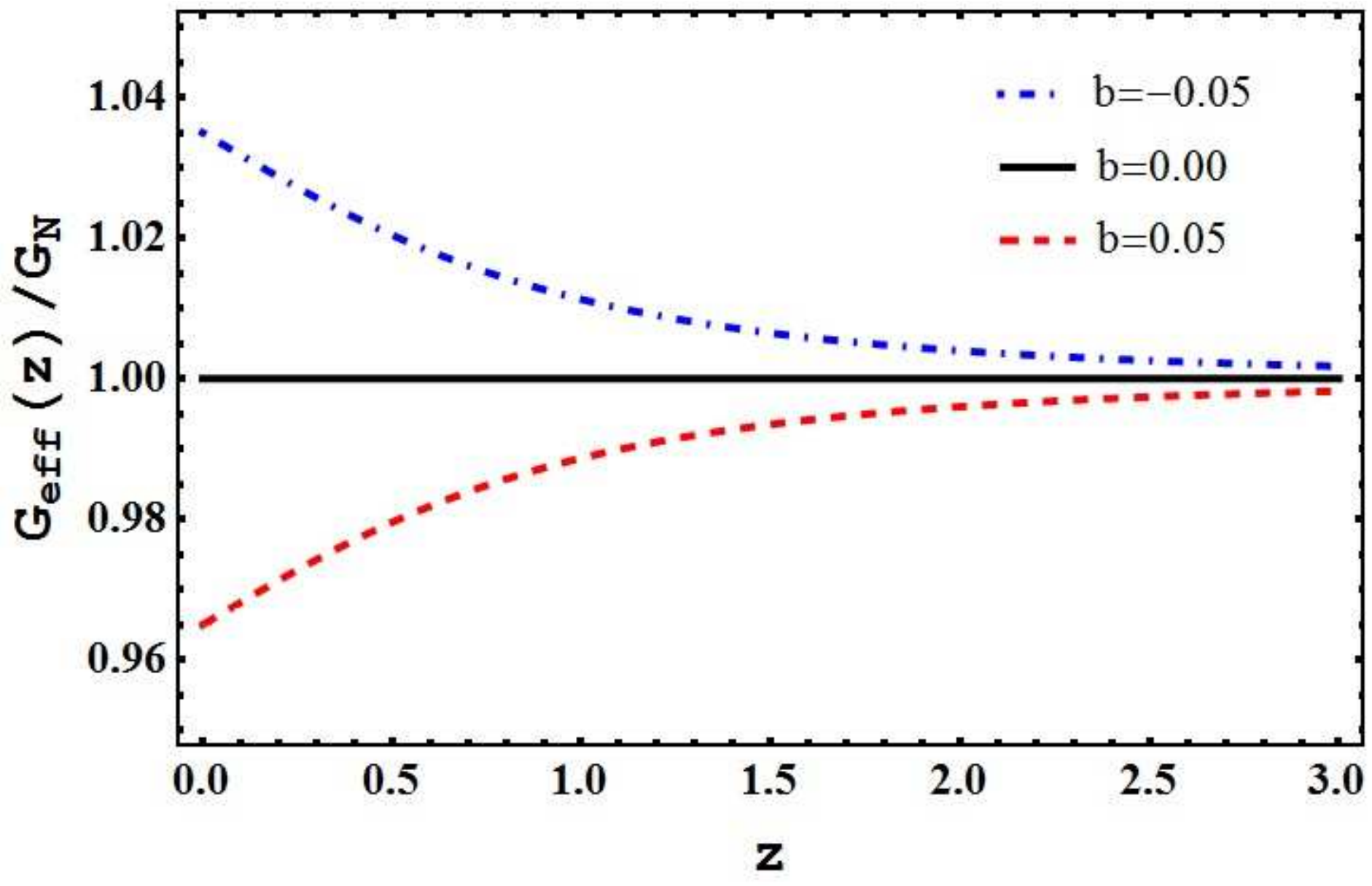}
 \caption{The evolution of $G_{\rm eff}/G_{N}$ in the case
of power-law $f(T)$ model.}
 \label{fig:Geff}
\end{figure}

In order to understand the differences of the $f(T) \propto
(-T)^{b}$ model from the concordance $\Lambda$ cosmology we plot in
Fig. (\ref{fig:Geff}) a comparison of the evolution of $G_{\rm
eff}/G_{\rm N}$. Notice, that the solid, dashed and the
dotted-dashed curves correspond to $b=0.00$ ($\Lambda$CDM), $0.05$ and
$-0.05$. We observe that at high redshifts $f(T) \propto (-T)^{b}$
tends to GR ($G_{\rm eff} \to G_{\rm N}$), but as we approach the
present time the ratio $G_{\rm eff}/G_{\rm N}$ starts to deviate
from unity. As an example, at $z=0$ the relative deviation from GR
is close to $\pm 4\%$ for $b=\pm 0.05$. We also find that a positive
value of $b$ implies that $G_{\rm eff}<G_{\rm N}$, while the
opposite holds for $b<0$.

The obvious connection
between $G_{\rm eff}$ and $b$ implies that the
free parameter $b$ should leave an imprint in 
the non-linear matter perturbations via
equation (\ref{matperteq3}). Indeed,
using equation (\ref{Geff1}) and changing the
variables from $t$ to $a(t)$ we obtain
\begin{eqnarray}\label{eqn:Nclusteing1}
&&\delta_{\rm
m}^{\prime\prime}+\left(\frac{3}{a}+\frac{E^{\prime}}{E}\right)\delta_{\rm
m}^{\prime} -\frac{4}{3}\frac{\delta_{\rm m}^{\prime
2}}{1+\delta_{\rm m}}- \frac{3\Omega_{\rm
m0}}{2a^5E^2}\times\\\nonumber &&(1-\frac{\Omega_{\rm F0}b}{E^2_{\rm
\Lambda}})\delta_{\rm m}(1+\delta_{\rm m})=0\;.
\end{eqnarray}

\begin{figure}
 \centering
 \includegraphics[width=8cm]{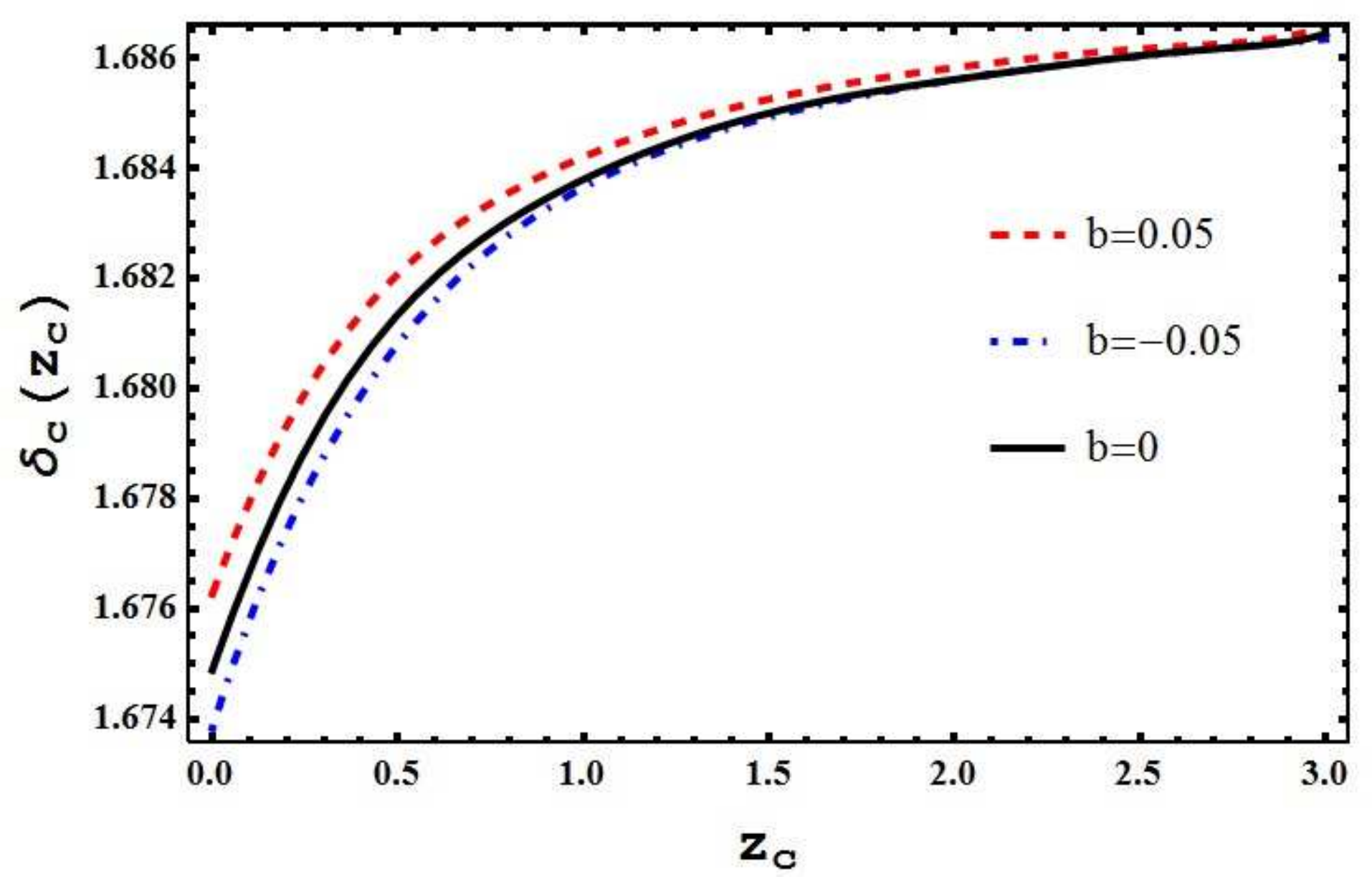}
 \caption{The critical overdensity $\delta_{\rm c}$ as a function of
the collapse redshift $z_{\rm c}$. The corresponding curves are
explained in the caption of Fig. (\ref{fig:background}).}
 \label{fig:deltac}
\end{figure}

Now in order to determine $\delta_{\rm c}$ and $\Delta_{\rm vir}$ we
follow the general approach of
\citep{Pace2010,Pace2012,Malekjani:2015pza,Pace:2014taa}. Specifically,
regarding $\delta_{\rm c}$ we utilize a two-step process. First, we
numerically solve equation (\ref{eqn:Nclusteing1}) between the epoch
$z_{\rm i}$ and the collapse redshift $z_{\rm c}$. As we have
already mentioned in the previous section concerning the value of
the initial scale factor of the universe $a_{\rm i}=1/(1+z_{\rm i})$
we use $10^{-4}$. Our attempt is to calculate the initial values
$\delta_{\rm mi}=\delta_{\rm m}(a_{\rm i})$ and
$\delta^{\prime}_{\rm mi}=\delta_{\rm mi}/a_{\rm i}$ for which the
collapse takes place at $a=a_{\rm c}$ such that $\delta_{\rm
m}(a_{\rm c}) \simeq 10^{7}$ \citep[see
also][]{Malekjani:2015pza,Nazari-Pooya:2016bra}. Second, we utilize
the values for $\delta_{\rm mi}$ and $\delta^{\prime}_{\rm mi}$
obtained in the first step as the initial conditions for the linear
equation (\ref{eqc1}) a numerical solution of which provides the
critical overdensity threshold above which structures collapse
$\delta_{c}\equiv \delta_{\rm m}(z=z_{\rm c})$. We remind the reader
that in the case of the Einstein-de Sitter model $\delta_{\rm c}$ is
strictly equal to 1.686. In Fig. (\ref{fig:deltac}), we show
$\delta_{\rm c}$ as a function of the collapse redshift $z_{\rm c}$
for the models explored here. We verify that $\delta_{\rm c}$
converges to the Einstein de-Sitter value at high redshifts, since
the matter component dominates the cosmic fluid. The $f(T)$ critical
overdensity starts to deviate from that of $\Lambda$CDM for $z \le
1.5$. In this redshift regime we observe that the critical
overdensity satisfies: $\delta_{\rm c}(z_{\rm c})>\delta_{\rm
c,\Lambda}(z_{\rm c})$ for $b=0.05$ and $\delta_{\rm c}(z_{\rm
c})<\delta_{\rm c,\Lambda}(z_{\rm c})$ in the case of $b=-0.05$.
This result is compatible with that of DE cosmologies  \citep[see][]{Pace2010,Devi:2010qp,Pace:2013pea,Nazari-Pooya:2016bra} .

Furthermore we apply the following fitting function \citep[see
also][]{Kitayama:1996ne,Weinberg:2002rd} to $\delta_{\rm c}$
 calculated in power law $f(T)$ gravity
\begin{equation}
\delta_{\rm c}(z)=\frac{3(12\pi)^{2/3}}{20}\Big(1+\beta\log_{\rm
10}{\Omega_{\rm m}(z)}\Big)\;,
\end{equation}
and obtain the constant coefficient $\beta$ in terms of parameter
$b$ as
\begin{eqnarray}
\beta=-0.04b+0.013
\end{eqnarray}

Another important quantity is the density contrast at virialization
which is defined as $\Delta_{\rm vir}=\xi(x/y)^3$, where $\xi$ is
the density contrast at the turnaround point, $x=a_{\rm c}/a_{\rm
ta}$ is the normalized scale factor with respect to the turn around
scale factor and $y$ is the ratio between virial radius and
turn-around radius, $y=R_{\rm vir}/R_{\rm ta}$ \citep{Wang1998}. It
is well known that for the Einstein de-Sitter model we have $(a_{\rm
c}/a_{\rm ta})_{\rm EdS} =(1+z_{\rm ta})/(1+z_{\rm c})= 2^{2/3}$,
$y=1/2$, $\xi=\left(\frac{3 \pi}{4}\right)^{2} \approx 5.6$ and thus
$\Delta_{\rm vir} \simeq 18\pi^{2}\simeq 178$. However, in DE
cosmologies the above quantities varies with the collapse redshift
\citep{Lahav1991,Wang1998,Mota2004,Horellou2005,Wang2005,Abramo2007,Basilakos:2006us,
Pace2010,Pace2012,Batista:2013oca,Pace:2013pea,Pace:2014taa,Malekjani:2015pza,Naderi2015}.

\begin{figure}
 \centering
\includegraphics[width=8cm]{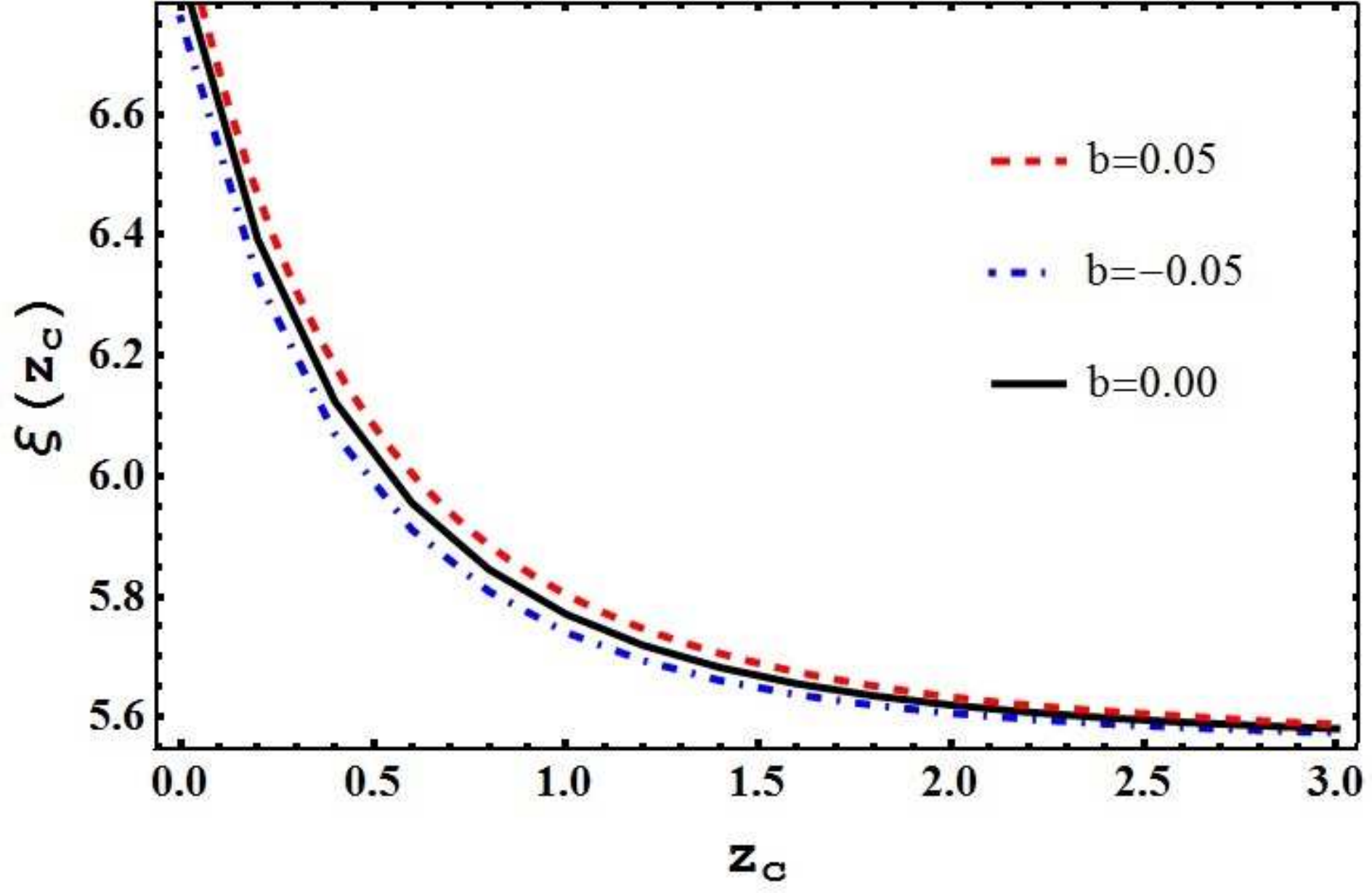}
\includegraphics[width=8cm]{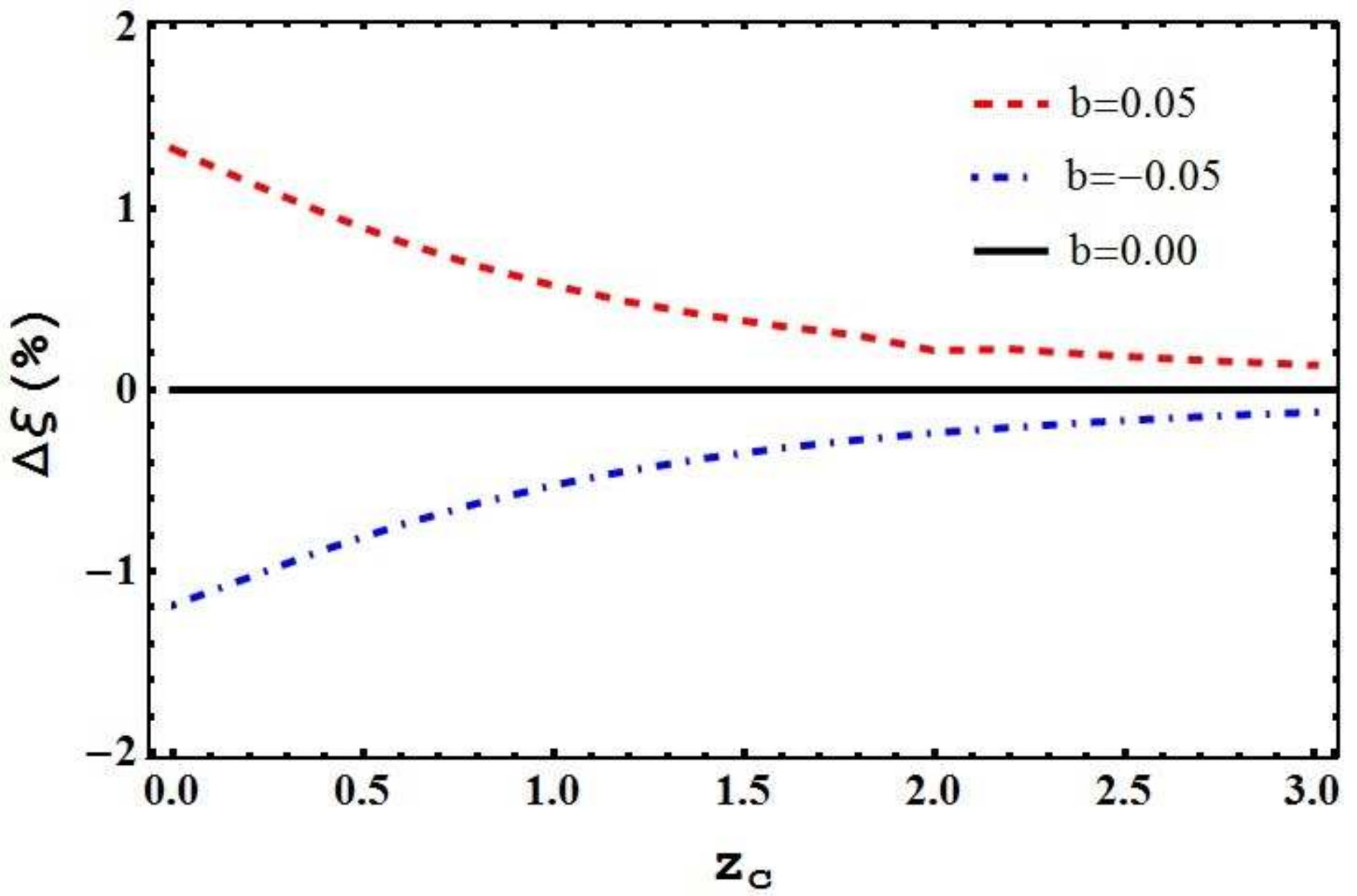}
 \caption{{\it Upper panel:}
The evolution of the overdensity $\xi$ at the turn around
point. {\it Lower panel:}
The fractional difference $\Delta \xi$
between the power law $f(T)$ model and the reference
$\Lambda$CDM model.
The lines correspond
to the same styles as in Fig. (\ref{fig:background}).}
 \label{fig:zeta}
\end{figure}

\begin{figure}
 \centering
\includegraphics[width=8cm]{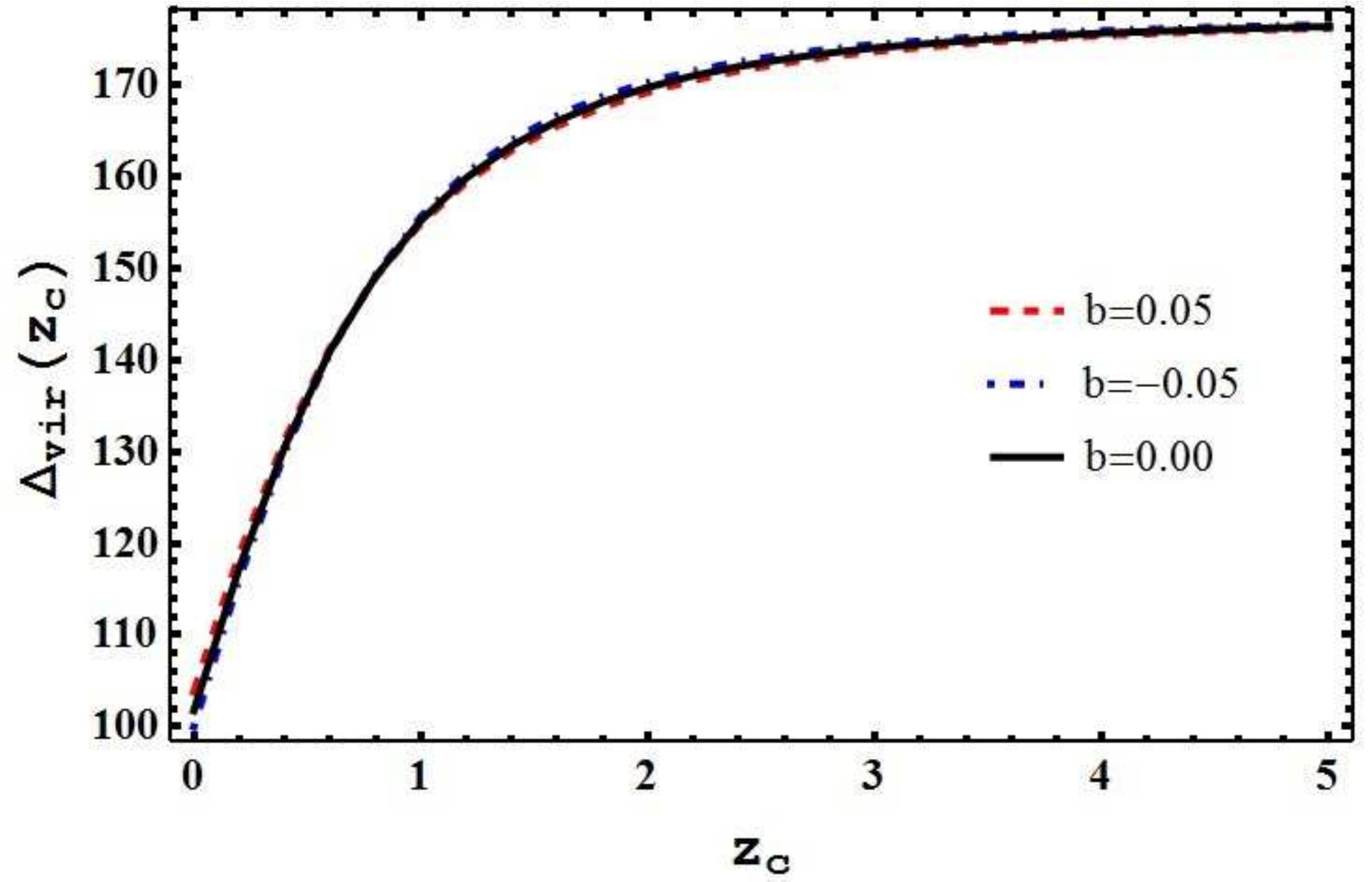}
\includegraphics[width=8cm]{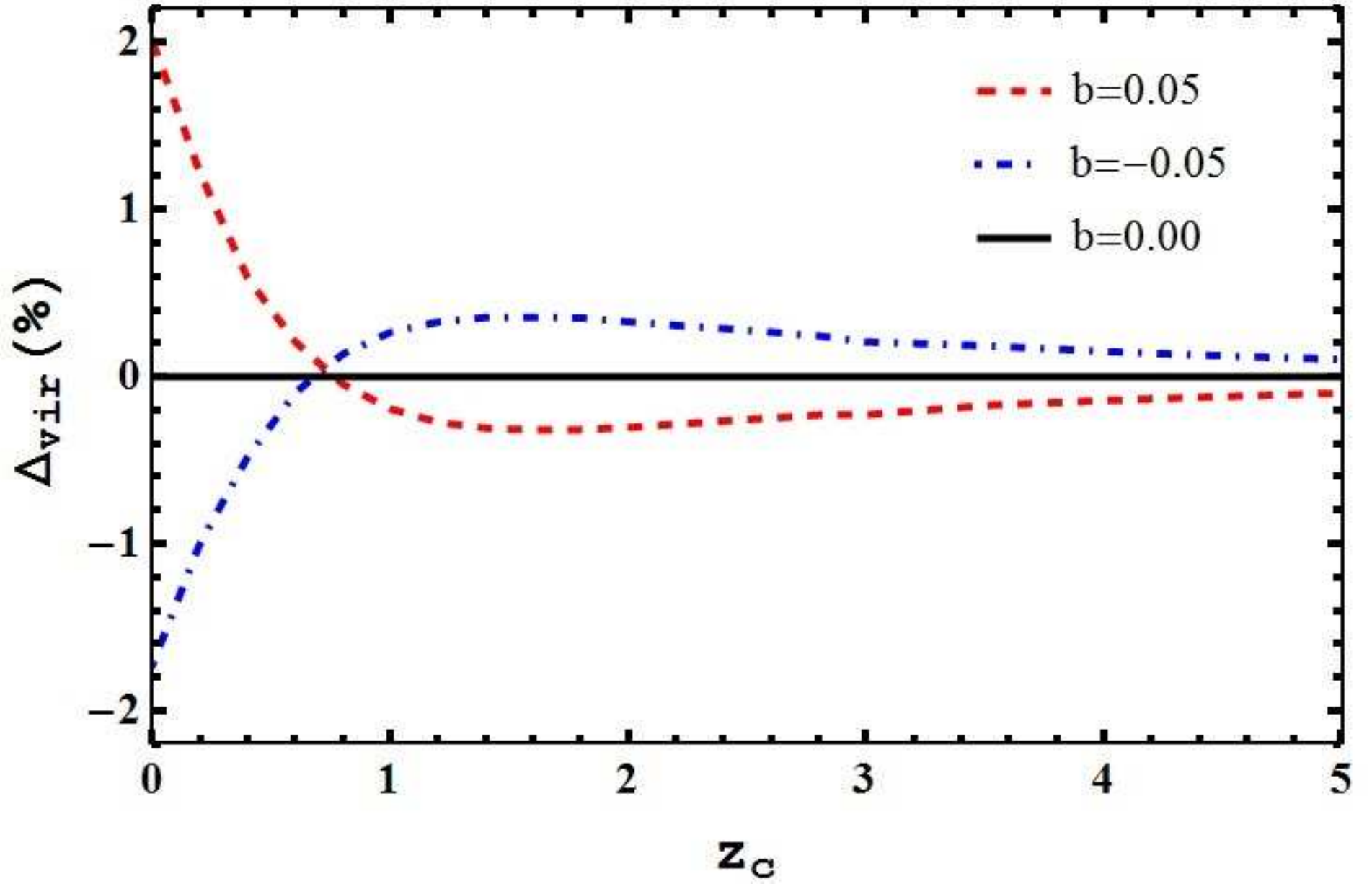}
 \caption{{\it Upper panel:} The virial overdensity
$\Delta_{\rm vir}$ as a function of the collapse redshift. {\it
Lower panel:} The fractional difference $\Delta_{\rm vir}(\%)$
versus $z_{\rm c}$.}
 \label{fig:deltavir}
\end{figure}

In the upper panel of Fig.(\ref{fig:zeta}) we plot the evolution
of the density contrast at turn around. Also, in the lower panel of
the same figure we present the relative difference deviation of the turn around density contrast $\xi(z_{c})$ for the power
law $f(T)$ model with respect to the $\Lambda$ solution
$\xi_{\Lambda}(z_{c})$. Obviously, the difference from the
$\Lambda$CDM case is small, namely at $z_{c} \sim 0$ we find $\sim
\pm 1.2\%$ for $b=\pm 0.05$. As expected, at very large redshifts
$\xi$ tends to the Einstein-de Sitter value ($\sim 5.6$). Moreover,
in the top panel of Fig. (\ref{fig:deltavir}) we provide
$\Delta_{\rm vir}$ as a function of $z_{\rm c}$ and in the bottom
panel of the same figure we show the behavior of $\Delta_{\rm
vir}(\%)=[(\Delta_{\rm vir}-\Delta_{\rm vir}^{\rm
\Lambda})/\Delta_{\rm vir}^{\Lambda}]\times 100$. At low redshifts
we find $\Delta_{\rm vir}(\%) \sim \pm 2\%$ for $b=\pm 0.05$.
Therefore, in the case of positive (negative) values of $b$ we
expect that the tendency for a large scale overdensity (candidate
structure) is to collapse in a more (less) bound system, with
respect to the $\Lambda$CDM cosmological model.

\section{Number of haloes}
\label{mass function}
\begin{figure*}
 \begin{center}
 \includegraphics[width=7cm]{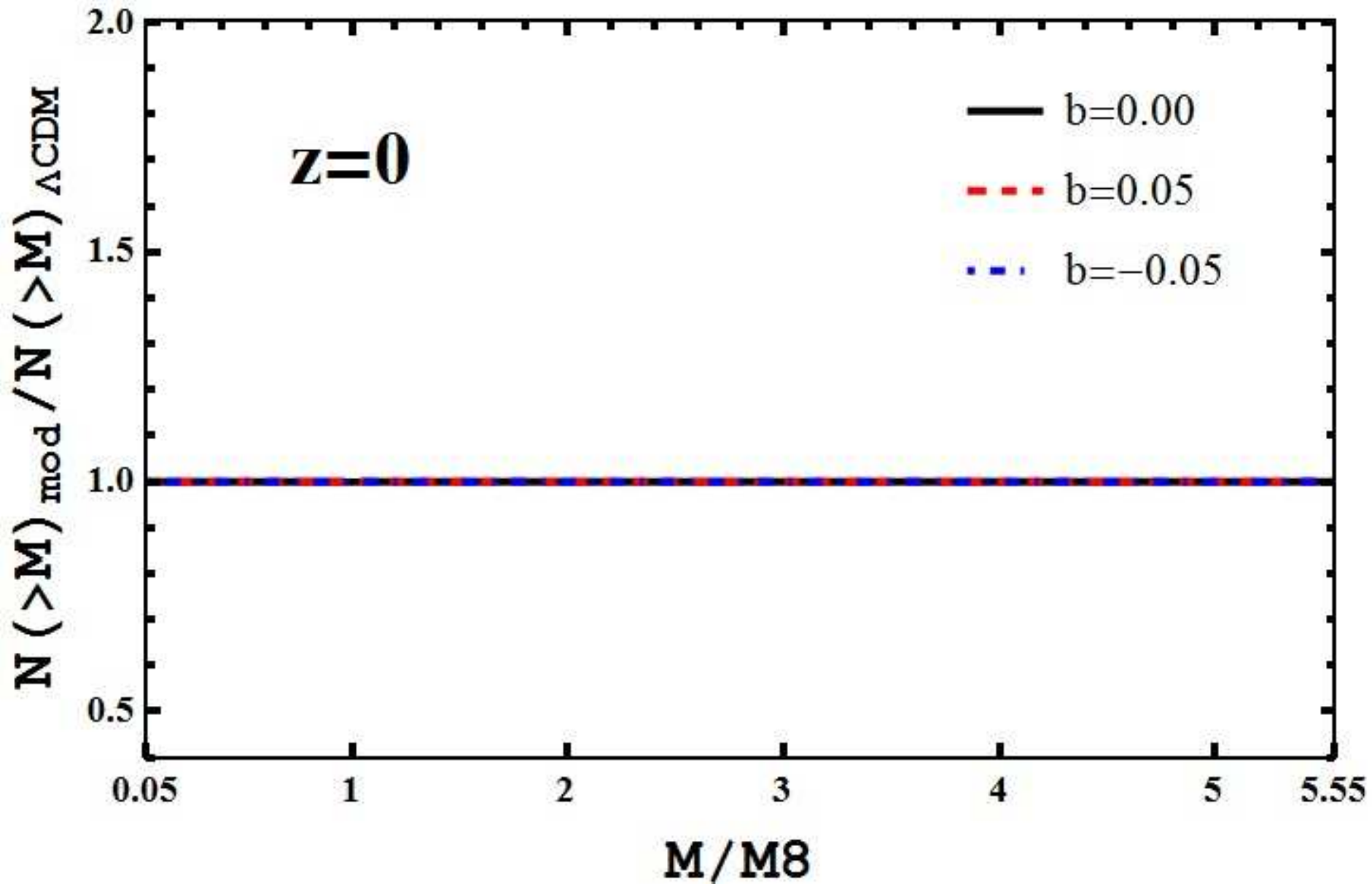}
\includegraphics[width=7cm]{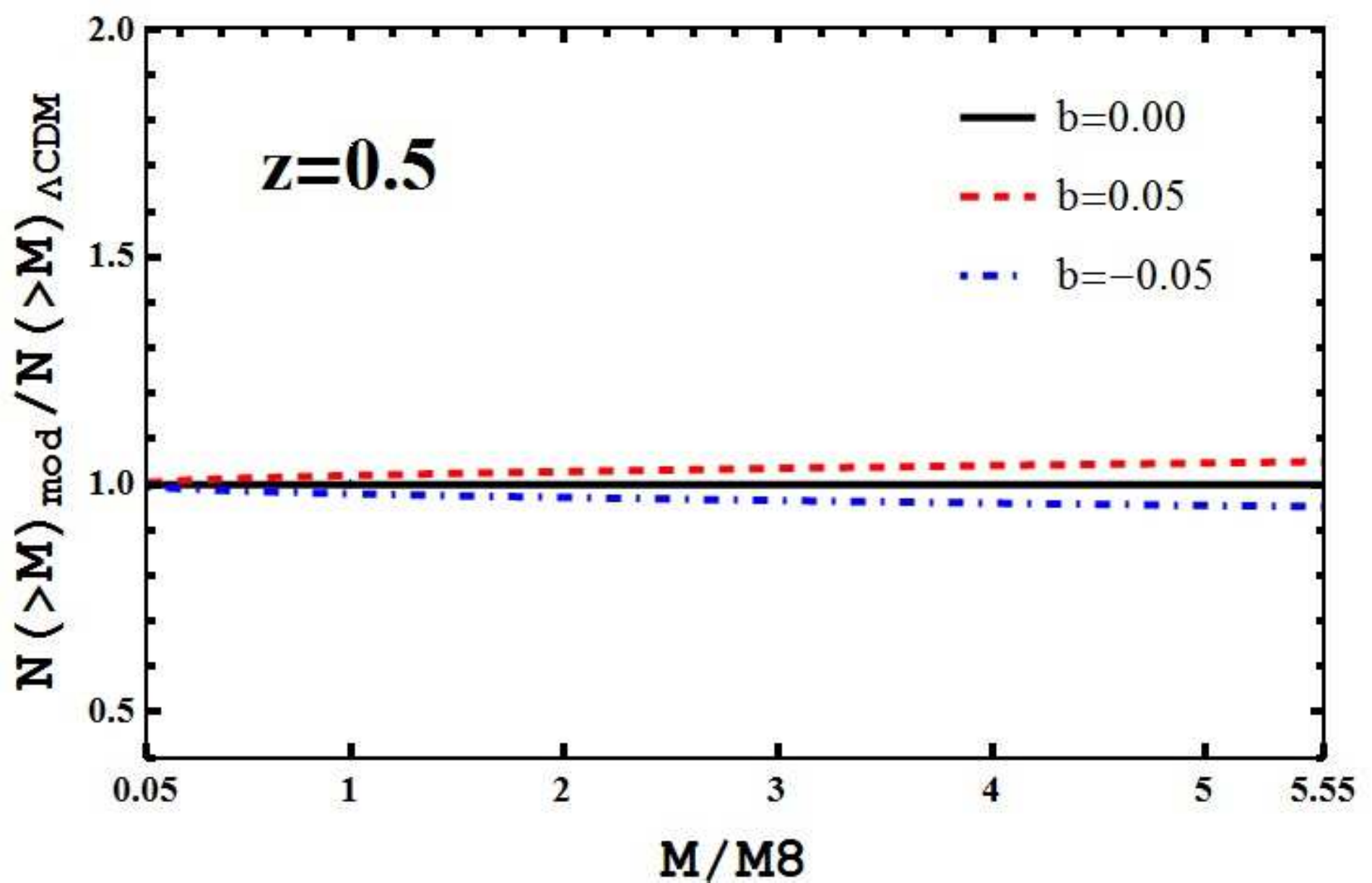}
\includegraphics[width=7cm]{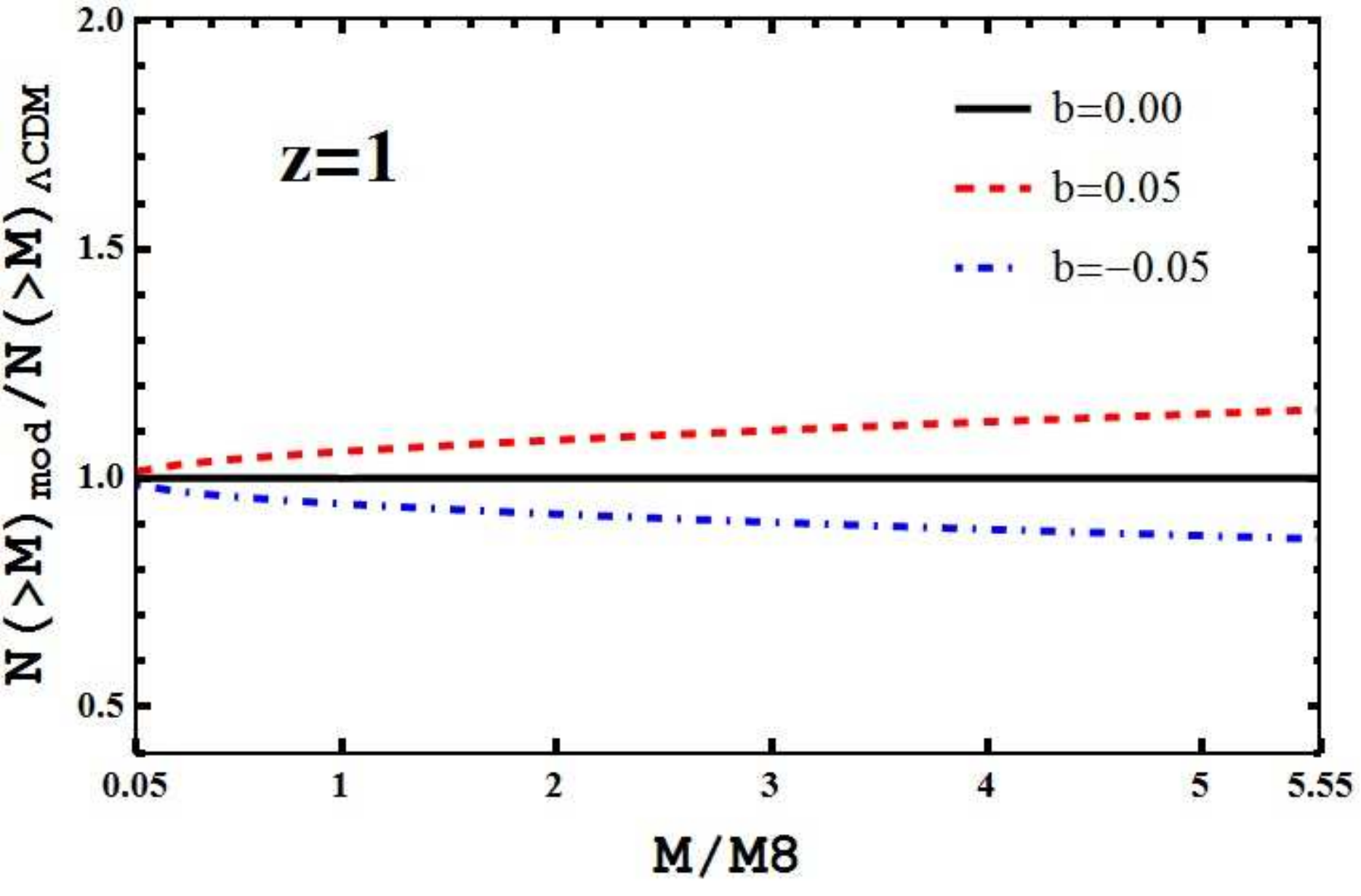}
 \includegraphics[width=7cm]{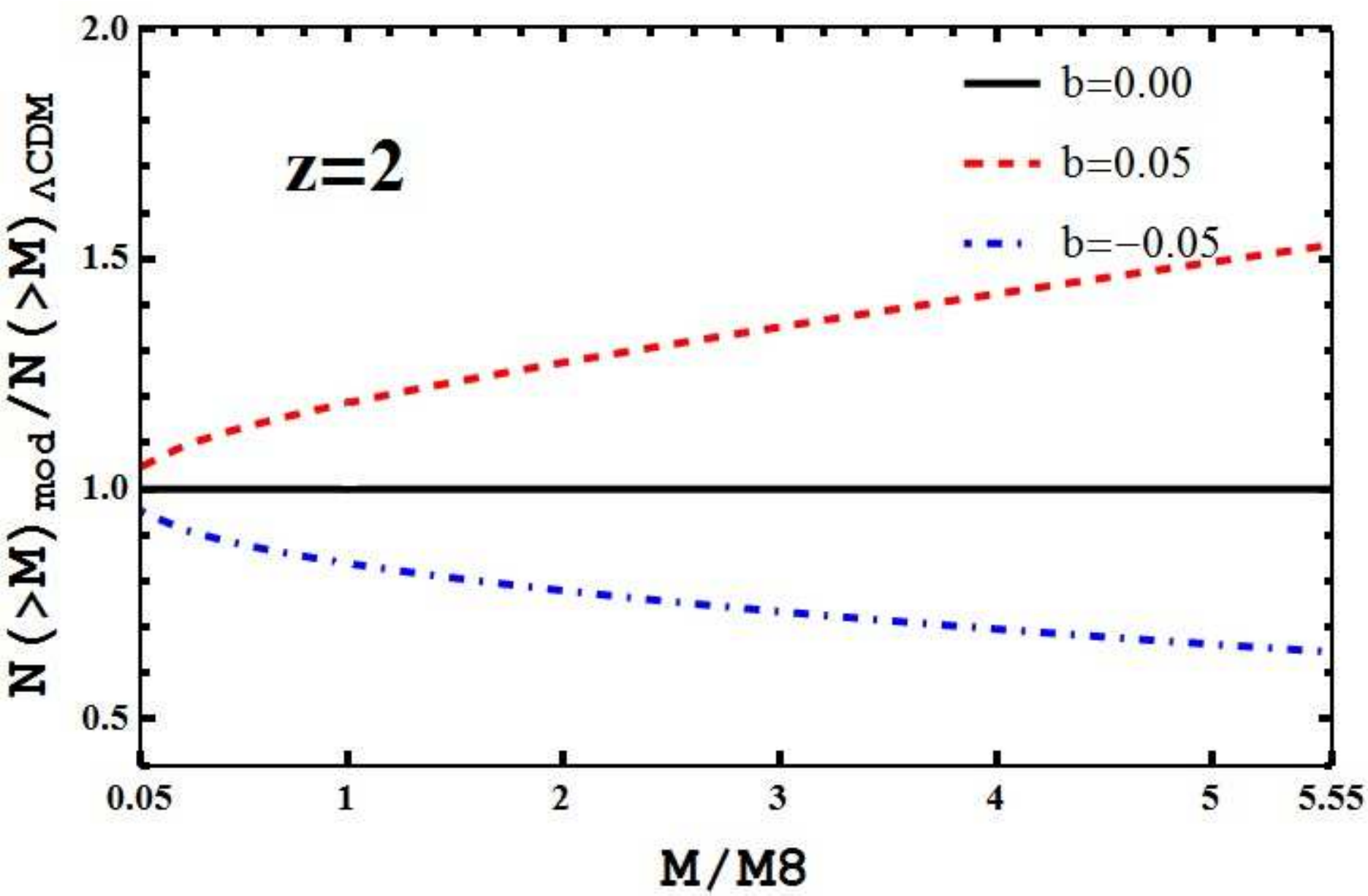}
 \caption{The expected ratio $N_{f(T)}/N_{\Lambda}$ as a function of $M/M_{8}$.Notice, we provide our results for different redshifts: $z=0$ (top left), $z=0.5$ (top right), $z=1.0$ (bottom left) and $z=2.0$. The style of curves
can be found in the caption of Fig.(\ref{fig:background}).}
  \label{fig:mass_fun1}
 \end{center}
\end{figure*}

In this section we compute the cluster-size halo number counts
within the framework of the cosmological models studied in this
article. Using the Press-Schechter formalism the abundance of
virialized haloes can be expressed in terms of their mass
\citep{Press1974}. The comoving number densities of virialized
haloes with masses in the range of $M$ and $M+dM$ is given by
\citep{Press1974,Bond1991}
\begin{eqnarray}
 \frac{dn(M,z)}{dM}=\frac{\rho_{m0}}{M}\frac{d\sigma^{-1}}{dM}f(\nu)\;,\label{eqn:PS1}
\end{eqnarray}
where $\nu(M,z)=\delta_{\rm c}/\sigma$, $\rho_{m0}=\Omega_{\rm
m0}\rho_{\rm cr,0}$ is the background density at the present time
and $\rho_{\rm cr,0} \simeq 2.775\times10^{11}$
$h^{2}M_{\odot}/Mpc^{3}$ is the corresponding critical density. In
the standard Press-Schechter approach the mass function is Gaussian
$f(\sigma)=\sqrt{2/\pi} \nu(\delta_c/\sigma) \exp(-\nu^2/2)$.
Notice, that $\sigma^2$ is the variance of the linear matter
perturbations
\begin{equation}\label{eq:sigma}
 \sigma^2(R,z)=\frac{D^{2}(z)}{2\pi^2}\int_0^{\infty}{k^2P(k)}W^2(kR)dk\;,
\end{equation}
where $R=(3M/4\pi\rho_{m0})^{1/3}$ is the radius of the spherical
region, $P(k)$ is the linear power spectrum and
$W(kR)=3[sin(kR)-kRcos(kR)])/(kR)^{3}$ is the Fourier transform of a
spherical top-hat profile.
 We utilize the cold dark matter (CDM) spectrum $P(k)=A k^{n}
T^{2}(\Omega_{\rm m0},k)$, with $T(\Omega_{\rm m0},k)$ the CDM
transfer function according to \citep{Eisenstein:1997ik} and
$n\simeq 0.96$, following the \cite{Planck:2015xua} results. In this framework, the rms matter fluctuations
is normalized at redshift $z=0$ so that for any cosmological model
one has \citep[for more detail see][]{Basilakos:2010fb}: $
\sigma^2(R,z)=\sigma^2_8(z) \frac{\Psi(\Omega_{\rm m0},
R)}{\Psi(\Omega_{\rm m0},
  R_8)}
$
with
$$
\Psi(\Omega_{\rm m0}, R)=\int_{0}^{\infty} k^{n+2} T^{2}(\Omega_{m0}, k) W^2(kR) dk
$$
and
$$
\sigma_8(z)=\sigma_8(0) D(z)\;,
$$
where $\sigma_{8}(0) [\equiv \sigma_8]$ the rms mass fluctuation on
$R_{8}=8 h^{-1}$ Mpc scales at redshift $z=0$. Concerning the value
of $\sigma_{8}$ we have set it to $\simeq 0.815$ based on the Planck
2015 results \citep{Ade:2015yua}. It is worth noting that the
Gaussian mass-function has a well known caveat, namely it
over-predicts/under-predicts the number of low/high mass halos at
the present epoch
\citep{Sheth1999,Jenkins:2000bv,Sheth2002,Lima:2004np}. In order to
avoid this problem in the present treatment we adopt the
Sheth-Torman (ST) mass function \citep{Sheth1999,Sheth2002}:
\begin{equation}
 f(\nu)= 0.2709\sqrt{\frac{2}{\pi}}\left(1+1.1096 \nu^{0.6}\right)
 \exp{\left(-\frac{0.707\nu^2}{2}\right)}\;.
 \label{eq:multiplicity_ST}
\end{equation}
Now given the determined mass
range, say $M_1\le M\le M_2$ we can derive the halo
number counts, $N(z)$ via the integration of the expected differential
halo mass function as
\begin{equation}\label{eq:ndensity}
N(>M,z)=
\int_{M}^{M_{2}}\frac{dn(z)}{dM^{\prime}}dM^{\prime}\;.
\end{equation}
In Fig.(\ref{fig:mass_fun1}), we display the expected ratio
$N_{f(T)}/N_{\Lambda}$ as a function of $M/M_{8}$ and above a
limiting halo mass, which is $M_{1} \equiv 10^{13}h^{-1}M_{\odot}$.
Concerning the upper mass limit we have set it to $M_{2} \equiv
10^{15}h^{-1}M_{\odot}$. We remind the reader that $M_{8}=6 \times
10^{14}\Omega_{\rm m0}h^{-1} M_{\odot}$ mass inside the radius of
$R_{8}=8 h^{-1}Mpc$ \citep{Abramo2007}.
Also the panels in Fig.~(\ref{fig:mass_fun1}) correspond to
different redshifts, namely $z=0$ (top left panel), $z=0.5$ (top
right panel), $z=1.0$ (bottom left panel) and $z=2.0$ (bottom right
panel). The results indicate that the number variation of the
differences between the $f(T)$ power law model and $\Lambda$
cosmology model is affected by variations in the value of $z$.
Considering $b=-0.05$ (or $b=0.05$) we find that significant model
differences should be expected for $z\gtrsim 1$, with the $f(T)$
model abundance predictions being always less (or more) than those
of the corresponding $\Lambda$ cosmology. In particular, at $z=1$
the $f(T)$ model with $b=0.05$ ($b=-0.05$) has roughly $1\%$ ($2\%$)
more (less) haloes than the standard $\Lambda$CDM model at the
low-mass tail $M/M_8=0.05$. Obviously, as we approach the high mass
haloes (see for example $M/M_8=5.55$) the corresponding differences
become more severe. Indeed, we observe that the $f(T)$ model with
$b=0.05$ ($b=-0.05$) produces $\sim 15\%$ ($\sim 12\%$) more (less)
haloes with respect to those of $\Lambda$CDM. Furthermore, the
deviation between $f(T)$ and $\Lambda$CDM models becomes even higher
at $z=2$. Specifically, for the low-mass tail $M/M_8=0.05$ we find
that the difference between $f(T)$ and $\Lambda$CDM can reach up to
$\pm \sim 5\%$ for $b=\pm 0.05$, while for the high-mass end
($M/M_8=5.55$) we show that the $f(T)$ gravity with $b=0.05$ (or
-0.05) predicts $\sim 52\%$ (or $\sim 36\%$) more (or less)
virialized haloes. We would like to point that 
the aforementioned predictions of the power law $f(T)$ model 
are similar to those of DE models (quintessence and phantom)
which adhere to GR \citep[see][]{Pace:2014taa}. We have expected 
such a similarity because in the case of $b<0$ (or $b>0$) the
power law $f(T)$ model is in the quintessence (or phantom) 
regime, namely
the effective EoS parameter obeys $w_{\rm de}>-1$ 
(or $w_{\rm de}<-1$) [see Fig. (\ref{fig:background})]. 

Although our analysis is self-consistent, in the sense that we compare
the expectations of $f(T) \propto (-T)^{b}$ model with respect to those of
the concordance cosmology using the same mass function, we want to 
investigate how
sensitive are the observational predictions to the different
mass functions fitting formulas.
For comparison, we use 
the mass function provided by \cite{Reed:2006rw}:

\begin{equation}
 f(\nu)= 0.2709\sqrt{\frac{2}{\pi}}\left(1+1.1096 \nu^{0.6}+0.2 G_1\right)
 \exp{\left(-\frac{0.763\nu^2}{2}\right)}\;,
 \label{eq:multiplicity_Reed1}
\end{equation}
where 
\begin{equation}
G_1=\exp{(-\frac{[\ln{\sigma}^{-1}-0.4]^2}{0.72})}\;.
 \label{eq:multiplicity_Reed2}
\end{equation}
We conclude that the difference between 
ST mass function and Reed et al. mass function is 
negligible at low mass tails and low-redshifts respectively. 
However, as we approach the high mass tail at $z=2$, we find 
$3\%-6\%$ differences between the two mass functions. Specificaly, 
for $b=0.05$ ($b=-0.05$) the mass function of \cite{Reed:2006rw} 
provides $\sim 6\%$ ($\sim 3\%$) more (less) haloes with 
respect to ST mass function. Overall, we verify that there are observational
signatures that can be used to differentiate the power law $f(T)$
gravity from the $\Lambda$CDM and possibly from a large class of DE
models \citep[see also][]{Basilakos:2010fb,Malekjani:2015pza}.

\section{conclusion}
\label{conclusion}
In this article, we have studied the spherical collapse model
(SCM)  and the
number counts of massive clusters beyond the concordance $\Lambda$
cosmology by utilizing the power law model for
the $f(T) \propto (-T)^{b}$ gravity.

First, at the level of the resulting cosmic expansion we have found
that the evolution of the main cosmological quantities are affected
by the power-law parameter, $b$. In particular, for $b<0$ we have
shown that the effective EoS parameter of the $f(T)$ gravity is in
the quintessence regime ( $w_{\rm de}>-1$), while it goes to phantom
($w_{\rm de}<-1$) in the case of $b>0$. Concerning the Hubble
parameter, we have found that the $f(T)\propto (-T)^{b}$ model is
close to that of the $\Lambda$CDM model (the relative difference can
reach up to $\sim 0.6\%$), as long as they are confronted with the
quoted set of observations.

Second we have investigated analytically and numerically the linear
and non-linear (via SCM) regimes of the matter perturbations in the
context of the current $f(T)$ gravity. In this case we have found
that the general behavior of the growth factor is similar to that of
the $\Lambda$CDM cosmological model, although the relative
difference is close to $1\%$ at high redshifts. 
We have showed that at low redshidts the linear growth 
of matter perturbations are suppressed due to the modifications of gravity while
at high redshifts the effect of modified gravity is less important.
Extending the $f(T)$
model in the non-linear phase of matter perturbations, we have
computed the well known SCM parameters, namely the linear
overdensity $\delta_{\rm c}$ and the virial overdensity $\Delta_{\rm
vir}$. We have showed that $\delta_{c}$ and $\Delta_{\rm vir}$ are
affected by the value of $b$. As expected both quantities tend to
those of Einsten-deSitter model at high redshifts.
Also, we have found that the predictions of SCM model 
in the power law $f(T)$ model are similar with those
DE models (quintessence or phantom) which adhere to GR 
\citep[for comparison see][]{Pace:2014taa}.

Finally, despite the fact that the $f(T) \propto (-T)^{b}$ model
closely reproduce the $\Lambda$CDM Hubble parameter, we have shown
that the $f(T)$ model can be differentiated from the reference
$\Lambda$ cosmology on the basis of
their number counts of cluster-size halos.
Indeed, using the Press-Schechter formalism in the framework of 
Sheth-Torman (ST) mass function \citep{Sheth1999,Sheth2002},
we have found
clear signs of difference, especially at $z\ge 1$,
with respect to the $\Lambda$CDM predictions.
Therefore, the power-law $f(T)$ gravity
model can be distinguished from the
$\Lambda$CDM and possibly from a large class of DE models,
including those of modified gravity.
 Also, using the mass function of Reed et al. \cite{Reed:2006rw} 
we found that the difference between the two mass functions 
is negligible at low mass tails and low-redshifts respectively. 
However, as we approach the high mass tail at $z=2$ 
we found that the relative difference lies in the interval $3\%-6\% $.
To this end, in the light of future cluster surveys
the methodology of cluster number counts appears to be very competitive
towards testing the nature of dark energy on cosmological scales.
\bibliographystyle{mnras}
\bibliography{ref}

\label{lastpage}

\end{document}